\documentclass[11pt,a4paper]{article}
\usepackage{geometry}
\usepackage{graphics}
\usepackage{setspace}
\usepackage[scaled=0.9]{helvet}
\usepackage{graphicx}
\usepackage{subfig}
\usepackage{amssymb}
\usepackage{rotating}
\usepackage{arydshln}
\usepackage{multicol}
\usepackage{abstract}
\usepackage{ucs}
\usepackage[utf8x,utf8]{inputenc}
\usepackage[T1]{fontenc}
\usepackage[swedish, english]{babel}
\usepackage{fancyhdr}

\begin{document}
\bibliographystyle{acm}
\pagestyle{fancy}
\cfoot{\thepage}
\renewcommand{\abstractname}{}

\title{\fontfamily{phv}\selectfont{\huge{\bfseries{Toward the estimation of background fluctuations under newly-observed signals in particle physics}}}}
\author{
{\fontfamily{ptm}\selectfont{\large{Federico Colecchia}}}\thanks{Email: federico.colecchia@brunel.ac.uk}\\
{\fontfamily{ptm}\selectfont{\large{{\it Brunel University London, Kingston Lane, Uxbridge, UB8 3PH, United Kingdom}}}}
}
\date{}
\maketitle
\begin{onecolabstract}
When the number of events associated with a signal process is estimated in particle physics, it is common practice to extrapolate background distributions from control regions to a predefined signal window. This allows accurate estimation of the expected, or average, number of background events under the signal. However, in general, the actual number of background events can deviate from the average due to fluctuations in the data. Such a difference can be sizable when compared to the number of signal events in the early stages of data analysis following the observation of a new particle, as well as in the analysis of rare decay channels. We report on the development of a data-driven technique that aims to estimate the actual, as opposed to the expected, number of background events in a predefined signal window. We discuss results on toy Monte Carlo data and provide a preliminary estimate of systematic uncertainty.
\end{onecolabstract}

\begin{multicols}{2}
{\bf Keywords:}
29.85.Fj; High Energy Physics; Particle Physics; Large Hadron Collider; LHC; background discrimination; mixture models; latent variable models; sampling; Gibbs sampler; Markov Chain Monte Carlo; Expectation Maximisation; Multiple Imputation; Data Augmentation.

\section{Introduction}
\label{intro}

The task of data analysis in particle physics often deals with data sets comprising collision events that contain the signature of a scattering of interest as well as background events that correspond to uninteresting processes mimicking the signal. When estimating the number of signal events, background probability density functions (PDFs) are often extrapolated from control regions to a predefined signal window and are subsequently used in template fits. However, this can only provide an estimate of the expected, or average, number of background events under the signal, and cannot take into account the effect of fluctuations. In practice, when the number of signal events, $S$, is much higher than the size of the typical fluctuations on the number of background events, $\sigma_B=\sqrt{\left<B\right>}$, $\left<B\right>$ being the expected number of background events in the signal window, the discrepancy between $B$ and $\left<B\right>$ can be neglected.

Nonetheless, when the number of signal events is low enough, the difference can be sizable. This can occur in the early stages of data analysis following the discovery of a new particle, or in the analysis of low-cross section processes. In such cases, the expected number of background events in the signal window can be a biased estimate of the actual number.

We report on the development of a data-driven technique that aims to estimate the actual number of background events under an observed signal, as opposed to the expected number. Our algorithm makes it possible to decompose an input mixture of signal and background events, e.g. a collection of events that pass all selection criteria corresponding to the end-point of a given analysis. This allows the shape of the background PDF to be estimated from the data, thereby taking into account the effect of statistical fluctuations. The development of this technique was influenced by a number of statistical methods, most notably the Gibbs Sampler \cite{geman} for mixture model decomposition, Expectation Maximisation \cite{EM}, and Data Augmentation \cite{DA}.

\section{The algorithm}
\label{algo}

The algorithm that we use to decompose the input mixture of signal and background events is related to a method that we have proposed with reference to a different application to data analysis at high-luminosity hadron colliders \cite{gibbshep2, gibbshep}.

The PDF of the underlying statistical model has the form $F=\alpha_0 f_0(x) + \alpha_1 f_1(x)$, where $\alpha_0$ and $\alpha_1$ are the fractions of background and signal events in the input data set, respectively, with $\alpha_0 + \alpha_1=1$, and where $f_0$ ($f_1$) is the background (signal) PDF. In the context of this study, the variable $x$ is interpreted as the invariant mass of a set of final state particles.

A notable feature of our approach, when compared to classical mixture models where predefined subpopulation PDF shapes are typically enforced a priori, is the nonparametric definition of the subpopulation PDFs, $f_j$. At every iteration of the algorithm, individual events are mapped to signal or background on a probabilistic basis, and the estimate $\varphi_j$ of the subpopulation PDF $f_j$ at that iteration is obtained by means of spline interpolation\footnote{
We have used the alglib C++ library \cite{alglib} with this implementation of the algorithm.
} of the histograms of $x$ corresponding to those events that are mapped to signal or background at that iteration. 
This allows the algorithm to estimate generic deviations of the PDF shapes from the corresponding control sample templates due to fluctuations in the data. The shapes of the signal and background distributions in the data set analysed are ultimately estimated as splined histograms averaged over a predefined number of iterations. 

The pseudocode of the algorithm is given below, subscripts ``sig'' and ``bkg'' relating to signal and background, respectively. The value of quantity $v$ at iteration $t$ is denoted by $v^{(t)}$ throughout.

\begin{enumerate}
\item {\bf Initialization:} Set $\alpha_{bkg}=\alpha^{(0)}_{bkg}=\alpha_{sig}=\alpha^{(0)}_{sig}=0.5$, where $\alpha_{bkg}=\alpha_0$ and $\alpha_{sig}=\alpha_1=1-\alpha_{bkg}$. Initial estimates $\varphi^{(0)}_j$ of the subpopulation PDFs $f_j$, $j=0,1$, are given by splined one-dimensional histograms of $x$ obtained from high-statistics control samples.
\item {\bf Iteration $t$:}
\begin{enumerate}
\item Generate $z_{ij}^{(t)}$ for all events $i$ and distributions $j$ according to $P(z_{ij}^{(t)}=1 | \alpha_j^{(t-1)}, \varphi_j^{(0)},x_i) = \frac{\alpha_j^{(t-1)}\varphi_j^{(0)}(x_i)}{\alpha_0^{(t-1)}\varphi_0^{(0)}(x_i)+\alpha_1^{(t-1)}\varphi_1^{(0)}(x_i)}$. Both the nonparametric treatment of the PDFs and the use of $\varphi_j^{(0)}$ instead of $\varphi_j^{(t-1)}$ to map individual events to signal or background distinguish this implementation from the classical Gibbs sampler for mixture models.
\item Set $\alpha_j^{(t)}=\sum_{i=1}^N z_{ij}^{(t-1)}/N$, $j=0,1$.
\end{enumerate}
\end{enumerate}

We used a total number of 6,000 iterations, and averaged the PDF estimates, $\varphi_j$, over the last 4,000. These settings allowed the algorithm to reach convergence in all runs performed in this study, and no significant difference in the results was observed by changing them.

A more detailed description of this implementation of the algorithm can be found in \cite{gibbshep5}. The execution time was $\sim50~\mbox{s}$ per run on the data sets analysed using a 2~GHz Intel Processor with 1~GB RAM, which we consider reasonable for offline use. 

\section{Results}
\label{results}

We illustrate this technique on a toy Monte Carlo data set obtained superimposing a gaussian signal with a first-order polyomial background. In the following, we will interpret the signal distribution as an invariant mass distribution corresponding to
the decay of a particle with mass $m=125~\mbox{GeV/c}^2$ and width $1~\mbox{GeV/c}^2$.
We superimposed $S=200$ signal events to a total of 4,200 background events in the region $115~\mbox{GeV/c}^2<m<135~\mbox{GeV/
c}^2$, corresponding to an average of $\left<B\right>=1,600$ background events in the signal region, which is defined by $120~\mbox{GeV/c}
^2<m<130~\mbox{GeV/c}^2$. 

Due to statistical fluctuations in the data, different samples correspond to different numbers of background events in the signal window. In this study, the standard deviation on the number of background events with $120~\mbox{GeV/c}^2<m<130~\mbox{GeV/c}^2$ is $\sigma_B=\sqrt{\left<B\right>}=40$ events, which is sizable when compared to the number of signal events generated, $S=200$. This illustrative scenario is not dissimilar from the early stages of data analysis following the observation of a Higgs boson in the $\gamma\gamma$ final state at the Large Hadron Collider (LHC) at CERN \cite{higgs_LHC_2012_ATLAS, higgs_LHC_2012_CMS}.

High-statistics control samples were generated corresponding to 30,000 signal and 30,000 background events, and were used to obtain initial conditions on the signal and background PDF shapes. The function of the algorithm is essentially to iteratively refine those initial conditions based on the data, thereby taking into account the effect of statistical fluctuations. As a consistency check, the estimated fraction of background events in the input data set, $\hat{\alpha}_0$, was found to be in agreement with the true value within 2\% in all runs used in this study.

\begin{figure*}
\centering
\subfloat[]{
\includegraphics[scale=0.48]{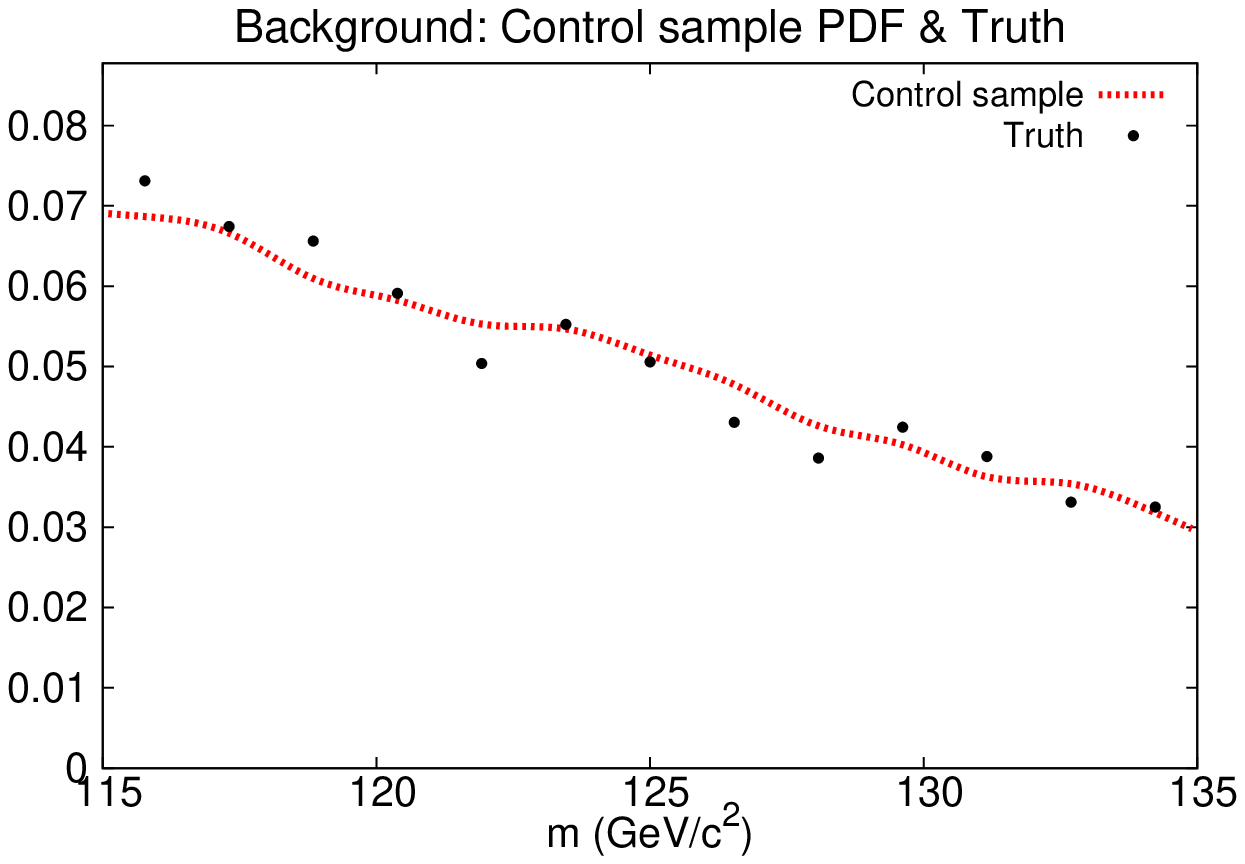}
}
\subfloat[]{
\includegraphics[scale=0.48]{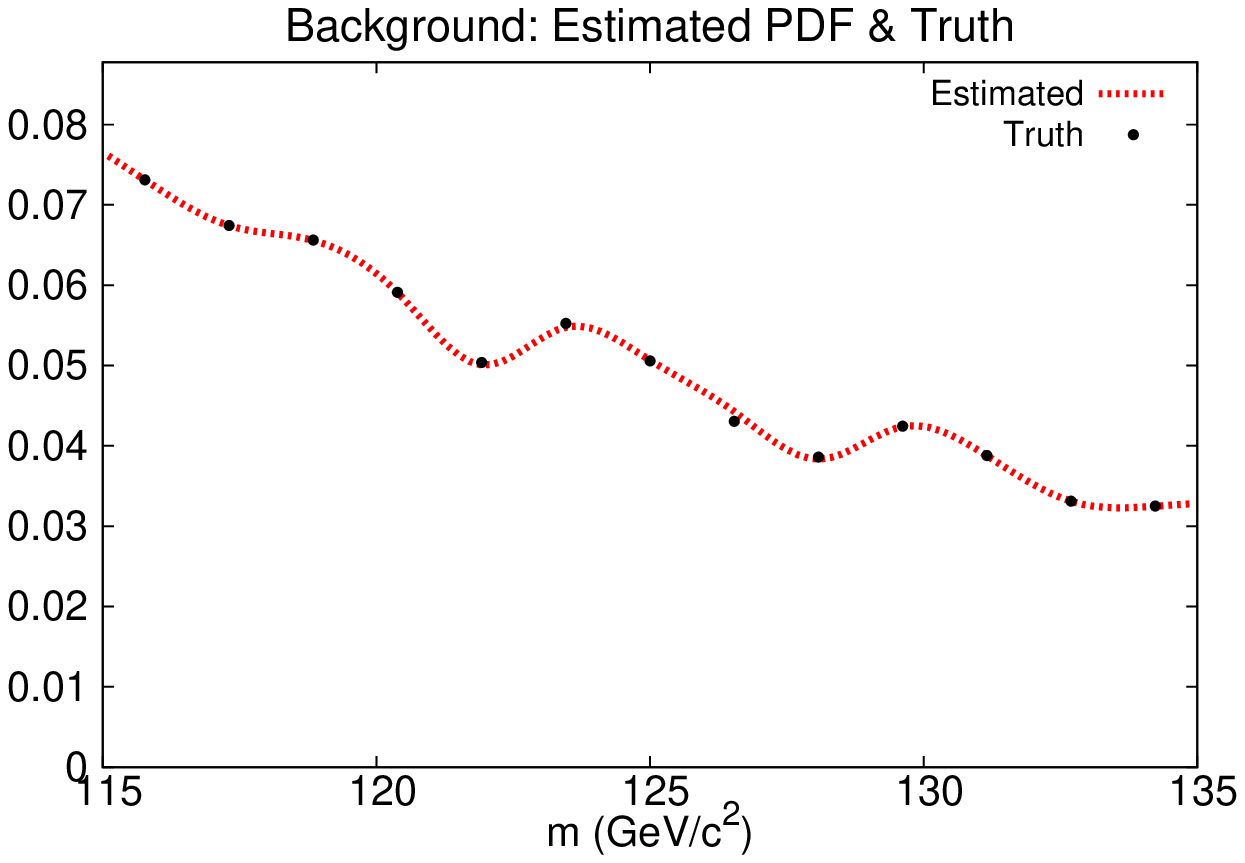}
}\\
\subfloat[]{
\includegraphics[scale=0.48]{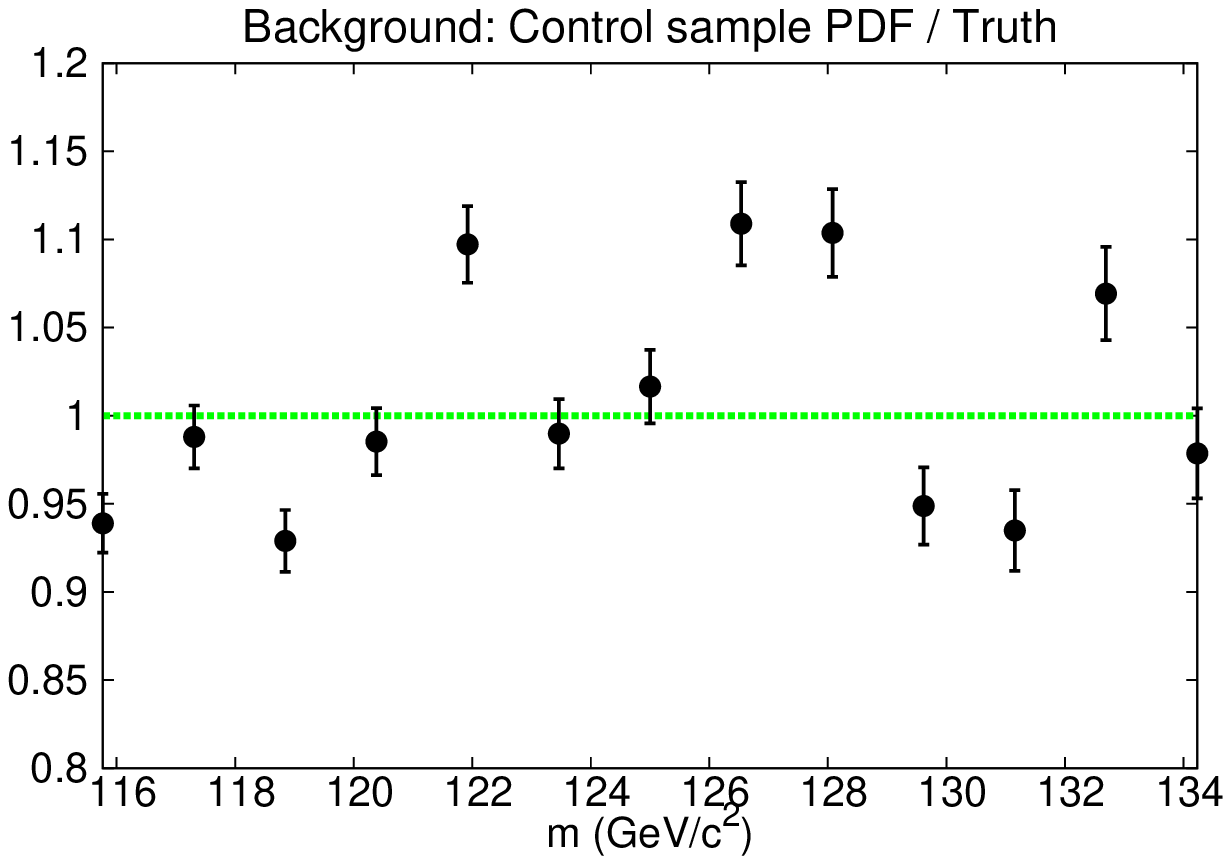}
}
\subfloat[]{
\includegraphics[scale=0.48]{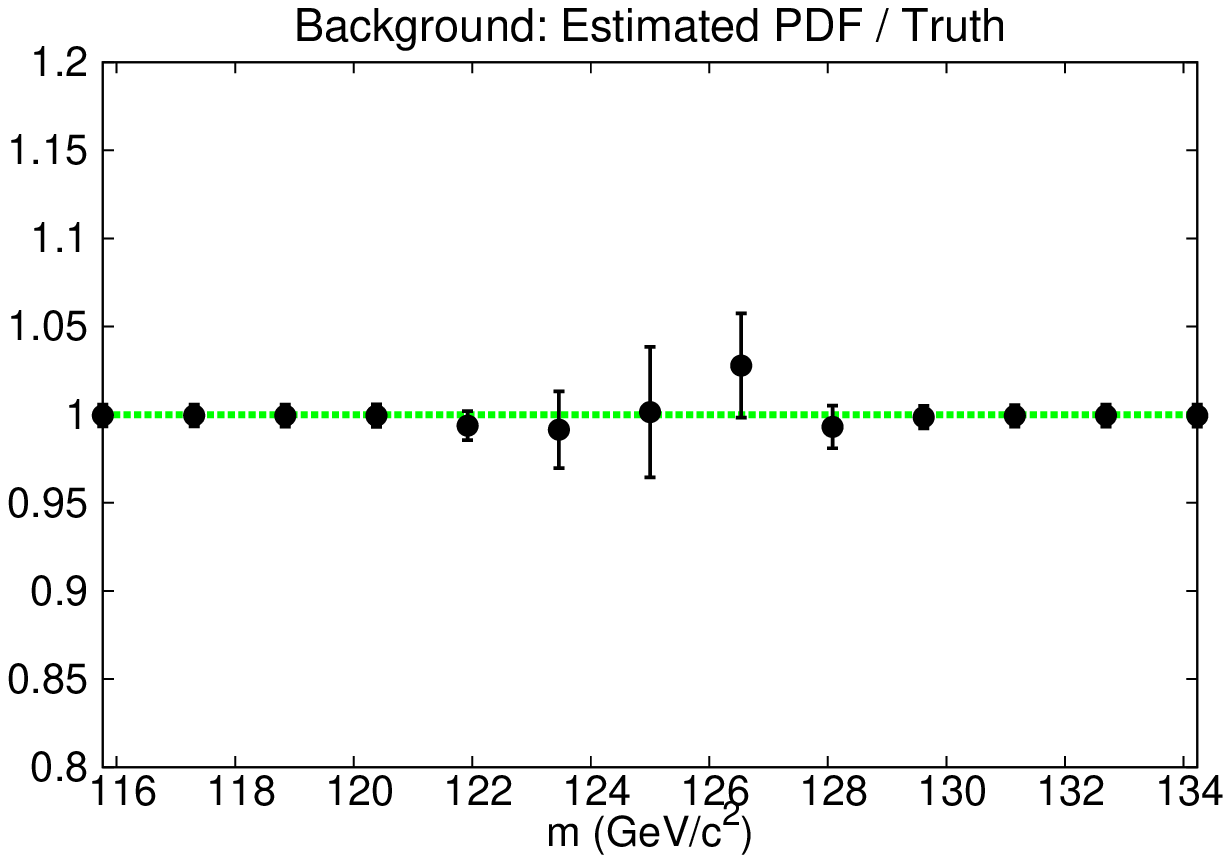}
}
\caption[]{
(a) True background distribution (points) superimposed with the PDF obtained from the high-statistics control sample (curve). (b) The same true background distribution (points) superimposed with the background PDF estimated from the data using the algorithm (curve). (c) Ratio between the background PDF obtained from the control sample and the true distribution. (d) Ratio between the background PDF estimated using the algorithm and the true distribution.
}
\label{fig:PDF_avg}
\end{figure*}

The performance of the algorithm in terms of estimating the shape of the background PDF in the data set analysed is illustrated in figure \ref{fig:PDF_avg}. Figure \ref{fig:PDF_avg} (a) displays the true background distribution (points) superimposed with the PDF obtained from the high-statistics control sample (curve). The discrepancies due to statistical fluctuations in the data are apparent. The points in figure \ref{fig:PDF_avg} (b) show the same true background distribution as in figure \ref{fig:PDF_avg} (a), but in this case the superimposed curve is the PDF estimated from the data using the algorithm, averaged over the last 4,000 iterations from a total of 6,000.

The ratio between the background control sample PDF and the true PDF is displayed in figure \ref{fig:PDF_avg} (c), which again highlights the effect of fluctuations. The corresponding ratio between estimated and true PDF is shown in figure \ref{fig:PDF_avg} (d), and shows a significantly-improved agreement.

It is worth recalling that, for the purpose of this study, what we are interested in is the shape of the background PDF. In fact, our objective is to estimate the actual number of background events under the signal as opposed to the expected number. The signal-related plots corresponding to figure \ref{fig:PDF_avg} showed good agreement between the estimated and the true distribution, and were used together with the estimated fraction of background events in the data in order to check the consistency of the results obtained using the algorithm.

The plots in figure \ref{fig:PDF_avg} refer to a run of the algorithm on a data set with $B=1,571$ background events in the signal region $120~\mbox{GeV/c}^2<m<130~\mbox{GeV/c}^2$. The corresponding number of events estimated with that run of the algorithm was $\hat{B}=1586.5$. 

The algorithm was also run on multiple toy Monte Carlo data sets, corresponding to different numbers of background events in the signal window. Our preliminary estimate of the uncertainty on $\hat{B}$, i.e. on the estimated number of background events under the signal, is $\sim50$ events. Work is underway to reduce this uncertainty below the size of typical background fluctuations in the data, $\sigma_B=\sqrt{\left<B\right>} = 40$ events. Our studies suggest that the uncertainty on $\hat{B}$ is dominated by the uncertainty on the estimated fraction of background events in the data set, $\hat{\alpha}_0$. In fact, when the algorithm is run with $\alpha_0$ kept fixed at the corresponding true value, the uncertainty on $\hat{B}$ drops from 50 to 12 events.

The results obtained running the algorithm on the different input data sets are summarised in table \ref{tab:syst_toy}, where $B_{gen}$ denotes the true number of background events in the signal window at generation, $\hat{B}$ is the corresponding number estimated using the algorithm, and $\Delta B = \hat{B}-B_{gen}$. The quantities $\hat{B}^*$ and $\Delta B^*$ in the table have a similar meaning as $\hat{B}$ and $\Delta B$, but the values were obtained running the algorithm with $\alpha_0$ kept fixed at its true value. The average and standard deviation of $\hat{B}$ across the runs are referred to as $\left<B\right>$ ( $\left<B\right>^*$) and $\sigma_B$ ($\sigma_B^*$), respectively.

\end{multicols}

\begin{table}[ht]
\centering % used for centering table
\begin{tabular}{c c c c c c} % centered columns (6 columns)
\hline\hline %inserts double horizontal lines
Run & $B_{gen}$ & $\hat{B}$ & $\Delta B$ & $\hat{B^*}$ & $\Delta B^*$ \\ [0.5ex] % inserts table
\hline % inserts single horizontal line
1 & 1536 & 1618.2 & 82.2 & 1549.3 & 13.3 \\% [1ex] % [1ex] adds vertical space
2 & 1569 & 1645.0 & 76.0 & 1592.3 & 23.3 \\% [1ex] % [1ex] adds vertical space
3 & 1579 & 1615.2 & 36.2 & 1584.7 & 5.7 \\% [1ex] % [1ex] adds vertical space
4 & 1625 & 1637.7 & 12.7 & 1630.2 & 5.2 \\% [1ex] % [1ex] adds vertical space
5 & 1558 & 1579.7 & 21.7 & 1548.0 & -10.0 \\% [1ex] % [1ex] adds vertical space
6 & 1576 & 1602.5 & 26.5 & 1588.2 & 12.2 \\% [1ex] % [1ex] adds vertical space
7 & 1571 & 1586.5 & 15.5 & 1579.1 & 8.1 \\% [1ex] % [1ex] adds vertical space
8 & 1584 & 1628.6 & 44.6 & 1584.8 & 0.8 \\% [1ex] % [1ex] adds vertical space
9 & 1597 & 1664.1 & 67.1 & 1604.4 & 7.4 \\% [1ex] % [1ex] adds vertical space
10 & 1644 & 1621.9 & -22.1 & 1640.4 & -3.7 \\% [1ex] % [1ex] adds vertical space
11 & 1631 & 1688.9 & 57.9 & 1636.6 & 5.6 \\% [1ex] % [1ex] adds vertical space
12 & 1573 & 1661.4 & 88.4 & 1586.9 & 13.9 \\% [1ex] % [1ex] adds vertical space
13 & 1626 & 1655.6 & 29.6 & 1616.3 & -9.7 \\% [1ex] % [1ex] adds vertical space
14 & 1583 & 1641.2 & 58.2 & 1592.8 & 9.8 \\% [1ex] % [1ex] adds vertical space
15 & 1613 & 1663.2 & 50.2 & 1635.8 & 22.8 \\% [1ex] % [1ex] adds vertical space
16 & 1593 & 1663.7 & 70.7 & 1606.1 & 13.1 \\% [1ex] % [1ex] adds vertical space
17 & 1583 & 1604.9 & 21.9 & 1585.2 & 2.2 \\% [1ex] % [1ex] adds vertical space
18 & 1603 & 1646.8 & 43.8 & 1586.8 & -16.2 \\% [1ex] % [1ex] adds vertical space
19 & 1624 & 1667.8 & 43.8 & 1630.4 & 6.4 \\% [1ex] % [1ex] adds vertical space
20 & 1580 & 1604.5 & 24.5 & 1575.4 & -4.6 \\% [1ex] % [1ex] adds vertical space
\hline %inserts single line
& $\left<\Delta B\right>$ = 42.5 & $\sigma_B$ = 27.3 & $\left<\Delta B^*\right>$ = 5.3 & $\sigma_B^*$ = 10.4 & \\
\hline
\end{tabular}
\caption{
Results obtained running the algorithm on different toy Monte Carlo data sets. The quantities $B_{gen}$ and $\hat{B}$ refer to the true and to the estimated number of background events in the signal region, respectively, and $\Delta B = \hat{B}-B_{gen}$. The quantities $\hat{B}^*$ and $\Delta B^*$ have a similar meaning as $\hat{B}$ and $\Delta B$, but the values were obtained keeping $\alpha_0$ fixed at its true value. The average and standard deviation of $\hat{B}$ across the runs are represented by $\left<B\right>$ ( $\left<B\right>^*$) and $\sigma_B$ ($\sigma_B^*$), respectively.
}
\label{tab:syst_toy}
\end{table}

\begin{multicols}{2}

\section{Conclusions and outlook}
\label{concl}

We have reported on the development of a data-driven technique that aims to estimate the actual, as opposed to the expected, number of background events under an observed signal in particle physics. Established methods that rely on the extrapolation of background distributions from control regions to a predefined signal window allow a precise estimation of the expected, or average, number of background events under the signal. However, the actual number of background events in the signal window can deviate from the average due to statistical fluctuations in the data. Although the discrepancy is often negligible when compared to the number of signal events, it is not necessarily so in the early stages of data analysis following the discovery of a new particle, or more generally in the analysis of low-cross section processes.

We have described an algorithm that uses the data to estimate the shape of the background distribution in a predefined signal window, e.g. using the end-point of a given analysis i.e. a collection of events that pass all selection criteria. Control samples are used only to provide initial conditions for the background PDF, but the PDF shape is otherwise estimated directly from the same data set that contains the observed excess of signal events. We have discussed results on toy Monte Carlo data, with reference to an illustrative scenario that is not dissimilar from the early stages of data analysis following the discovery of a Higgs boson in the $\gamma\gamma$ channel.

We have provided a preliminary estimate of the uncertainty associated with the estimated number of background events in the signal window at the level of 50 events, out of a total average number $\left<B\right> = 1,600$. Although we consider these results encouraging, the uncertainty is still larger than the size of the typical background fluctuations in the data, which is given by $\sigma_B = \sqrt{\left<B\right>} = 40$ events. Work is underway to improve the performance of the algorithm in this respect. It should also be emphasised that, since the above uncertainty is expected to depend significantly on $B$, the assessment of the performance of this method will have to take into account the specifics of the analysis in question.

\section{Acknowledgments}
The author wishes to thank the High Energy Physics Group at Brunel University for a stimulating environment, and particularly Prof. Akram Khan, Prof. Peter Hobson and Dr. Paul Kyberd for fruitful conversations, as well as Dr. Ivan Reid for help on technical issues. Particular gratitude also goes to the High Energy Physics Group at University College London, especially to Prof. Jonathan Butterworth for his valuable comments. The author also wishes to thank Prof. Trevor Sweeting and Dr. Alexandros Beskos at the UCL Department of Statistical Science for fruitful discussions. Finally, particular gratitude goes to Prof. Carsten Peterson and to Prof. Leif Lönnblad at the Department of Theoretical Physics, Lund University.

%\section*{References}

\end{multicols}

\begin{thebibliography}{9}
\bibitem{geman}Geman S and Geman D 1984 {\it IEEE T. Pattern Anal.} {\bf 6}(6) 721-41
\bibitem{EM}Dempster A~P, Laird N~M and Rubin D~B 1977 {\it J. Roy. Statist. Soc. Ser.} B {\bf 39}(1):1-38
\bibitem{DA}Tanner M~A and Wong W~H 1987 {\it J. Amer. Statistical Assoc.} {\bf 82} (398):528-540
\bibitem{gibbshep2}Colecchia F 2013 {\it J. Phys.: Conf. Ser.} {\bf 410} 012028
\bibitem{gibbshep}Colecchia F 2012 {\it J. Phys.: Conf. Ser.} {\bf 368} 012031
\bibitem{gibbshep5}Colecchia F 2013 arXiv:1311.2300 [physics.data-an]
\bibitem{alglib}ALGLIB (www.alglib.net), Sergey Bochkanov
\bibitem{higgs_LHC_2012_ATLAS}The ATLAS Collaboration 2012 {\it Phys. Lett.} B {\bf 716}(1):1-29
\bibitem{higgs_LHC_2012_CMS}The CMS Collaboration 2012 {\it Phys. Lett.} B {\bf 716}:30-61
\end{thebibliography}
\end{document}